# Internal waves and vortices in satellite images

Amelia Carolina Sparavigna
Department of Applied Science and Technology
Politecnico di Torino, Italy

*Some recent papers proposed the use of the satellite images of Google Earth in teaching physics, in particular to see some behaviours of waves. Reflection, refraction, diffraction and interference are easy to be found in these satellite maps. Besides Google Earth, other sites exist, such as Earth Observatory or Earth Snapshot, suitable for illustrating the large-scale phenomena in atmosphere and oceans In this paper, we will see some examples for teaching surface and internal sea waves, and internal waves and the Kármán vortices in the atmosphere. Aim of this proposal is attracting the interest of students of engineering schools to the physics of waves.*

Some recent papers proposed the use of the satellite images, available by the Google Earth or Maps web-services, to show some real cases in teaching physics and other disciplines [1-3]. For geophysics, this use is obvious. With easy image processing methods, Google Maps can be enhanced to study the features of Earth surface textures [4-6]. For other disciplines, such as archaeology for instance, the use of satellites had been discussed in several papers [7].
The use of satellite maps provides several possibilities for teaching physics. Ref.[1] is approaching this use aiming to increase the appeal of physics to students. An example: to bring to life the stories of famous scientists by paying a "flying visit to their homes". The author suggests to visit Woolsthorpe Manor, the birthplace of Sir Isaac Newton. For studying waves, Ref.2 shows by means of several images of the sea near shorelines, how the satellites can record the behaviour of waves. Reflection, refraction, diffraction and interference are easy to be found, when the resolution is high enough. Another paper [3] is proposing the satellite images for an evaluation of the speed of the boats, applying a simplified approach to the Kelvin wave patterns.
Google Earth or Maps are free and easy to use. Unfortunately, they cannot help in teaching the physics of atmosphere and oceans. However, other sites exist, such as Earth Observatory or Earth Snapshot, that are suitable for this purpose [8,9]. The first is a NASA web site, the second is proposing satellite images from the collections owned by EOSNAP/Chelys. The images provide observations at large scales, and therefore, can be good for study and teach some cases different from those already proposed.
In this paper, we will see some examples suitable for visualizing surface and internal sea waves, not yet discussed in [2], with the Google Maps. Then the use of Earth Observatory or Snapshot is proposed for atmospheric internal waves and the Kármán vortices. Aim of these proposals is attracting the interest of students of engineering schools to the physics of waves.

**Transitional and shallow-water waves**
Instead of using a ripple tank, we can see all the wave phenomena in the satellite images of the sea, when Google or Bing Maps have an enough high resolution to observe the wavelets on the surface of water. Before discussing the waves near the shore and the internal waves, let me show a quite remarkable example of circular waves created by an object that struck vertically the water, at the same time the satellite captured the image (see Fig.1). A hunter of anomalies in Google Earth images observed this spot near the United Kingdom coastline [10]. Bing Maps reveal that there are no rocks, but what looks like a crater in the seabed.
Besides the examples in Ref.2, let me consider the waves near the shoreline, where they are transitional or shallow-water waves. We tell that we are in a condition of deep-water when the depth of water is greater than ½ of the wavelength. For a water depth between 1/2 and 1/20 the wavelength, we observe the transitional waves. Then we have the shallow water waves when the depth is 1/20 of the wavelength [11]. The shallow-water waves interact with the bottom of the sea and some of their characteristics are altered. The period remains constant but waves slow down. For this reason, their wavelengths shorten and wave heights increase. The wave steepness, which is the ratio of height to length, increases and the wave becomes more unstable.

An example of these waves is proposed in Fig.2. Images of this kind can be used in lessons such as [12] on the waves: quite interesting is the discussion of the wave energy, which can be proposed for students of engineering, attracted by the planning of coastal buildings and infrastructure.

**Internal waves**

What we are usually observing in the Google Maps are the surface waves at the interface between the air and sea. There are other waves we can try to find in the satellite images. These are the internal waves. They are occurring at the interface between two layers of sea or ocean waters having a differing density. Figure 3 is reporting some internal waves near the coast of Peru. The image is adapted from the Google Maps.

Internal waves can be originated from a temperature gradient in the water. Consequently, the density of water increases towards the bottom, creating a layered structure of different densities, from the seabed to the surface. The internal waves are then those disturbances that occur at the boundary between layers of different water density. Their wave heights can be quite large, such as the wavelengths. The origin of the waves can be the tidal movement, but also turbidity currents, wind or even passing ships. If the crests approach the surface, these crests affect the reflection of light from the water. Therefore, images of internal waves can be captured by satellites [13].

It seems that the internal oceanic waves were first seen from space by American astronauts on their way to the Moon. Since waves depend on gravity, a local mass oscillation can appear more easily at an internal interface than at the sea-surface. The difference in density between layers of water is smaller than the density difference between water and air, and then the energy required to have an internal waves is less than that required for surface waves of similar amplitude. [14]. As told in Ref.14, the waves influence the concentration of biological materials and natural oils, which are alternately dispersed and concentrated in a wave pattern. The natural materials change how the surface reflects light, revealing the underlying internal waves.

The internal waves exist in the atmosphere too. These waves and other perturbations can be observed when they produce some clouds. As an air mass is subjected to a perturbation, such as a wave, it oscillates up and down in the atmosphere. If there is enough moisture, clouds will form at the cool crests of the wave. In the descending part of the wave, the clouds disappear due to an adiabatic heating. Therefore, some clouded and clear bands are visible [15]. Google Maps are not suitable for the study of the atmosphere, because the maps are usually proposed without clouds, but we can use the images of the Earth Observatory or Earth Snapshot. There are several images of waves clouds. For example, that in Fig.5 [16], where we see the wave clouds over the Aral Sea. The image was captured by the Moderate Resolution Imaging Spectroradiometer (MODIS) on NASA's Aqua satellite. The clouds conform to the shape of the western shore. As Ref.16 is telling, the wave clouds typically form when a mountain or an island, or even another mass of air, forces air to oscillate.

**Wakes and vortices in the atmospheres**

Figure 5 shows what happens when the cloud waves are produced by islands [17]. Here, the islands are the Kerguelen Islands, also known as Desolation Island, in the southern Indian Ocean. As told in Ref.17, although these islands are hardly visible in satellite image, their location is easy to discern due to the presence of the wave clouds, having a form similar to that of the wake created by a ship. The islands are at the start of the wakes. We could apply the approach of Ref.3 for the wakes of boats to determine the motion of the atmosphere.

Impressive are other cloud patterns that islands can create in the atmosphere, as those shown by Fig.6 [18]. The vortices that we see are known as Kármán vortex streets, and are repeating patterns of swirling vortices caused in a flow of a fluid by bluff bodies. In this case, the bluff bodies are the Cape Verde Islands. Dust from the Sahara Desert blows off the west coast of Africa. When the wind encounters the island, a disturbance propagates downstream of the island as a row of vortices, which alternate their direction of rotation. We see them in the dust from the Sahara desert.

**Discussion**

A Kármán vortex street is then a repeating pattern of swirling vortices, named after Theodore von Kármán, engineer and fluid scientist, born in 1881 at Budapest. In 1930 he accepted the directorship of the Guggenheim Aeronautical Laboratory at the California Institute of Technology (GALCIT) and emigrated to the United States. In 1944 he and others affiliated with GALCIT founded the Jet Propulsion Laboratory [19].This very short reference to Kármán's life is an example of the use of satellite images as proposed in [1], to connect physics, society and history.

We often talk about phenomena named after a scientist, but we do not know much about her/his life. In my opinion, the satellite images can stimulate the analysis, besides of the natural phenomena of course, of the life of scientists and the role and influence in their social frameworks.

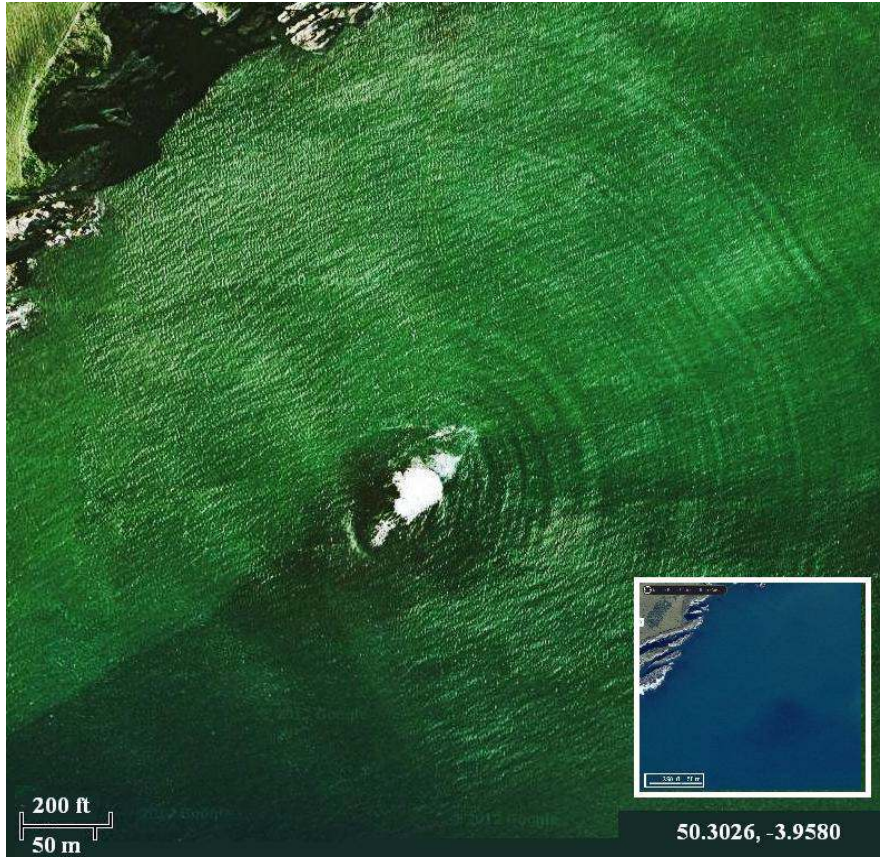

Fig.1: Circular waves created by an abject that fall vertically in the water (coordinates 50.3026,-3.9580, UK coastlines). Using Google Maps is possible to see the reflection of these circular waves on the coast. The inset image shows the same location as seen in the Bing Maps.

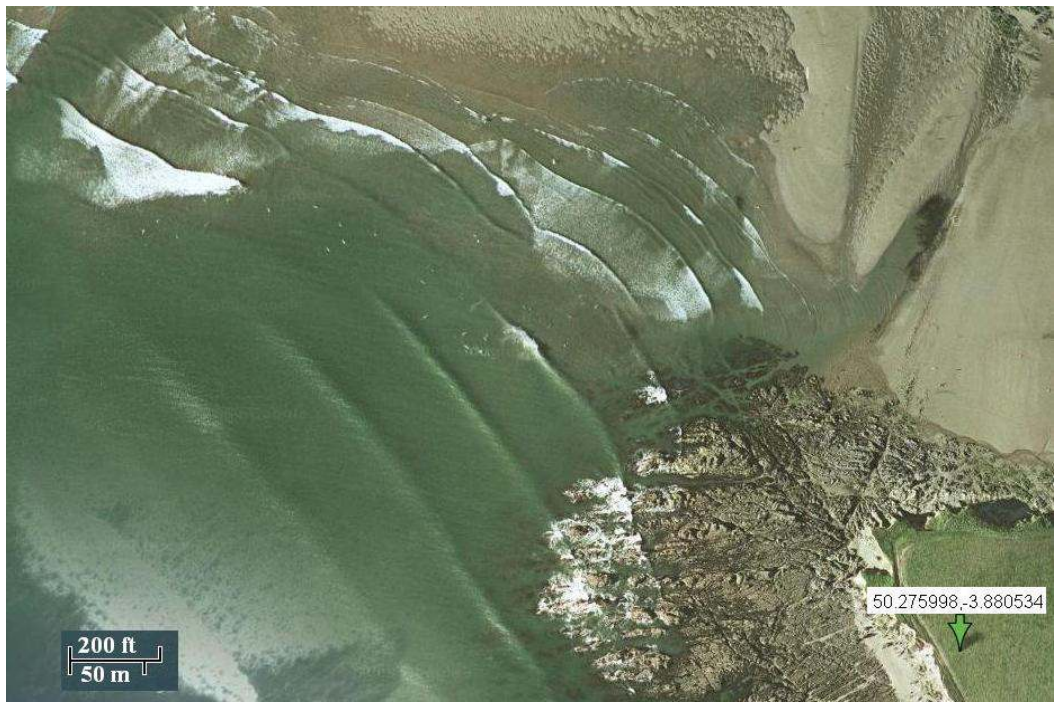

Fig.2 Transitional and shallow-water waves on the coastline of United Kingdom

(coordinates: 50.2760,-3.8805).

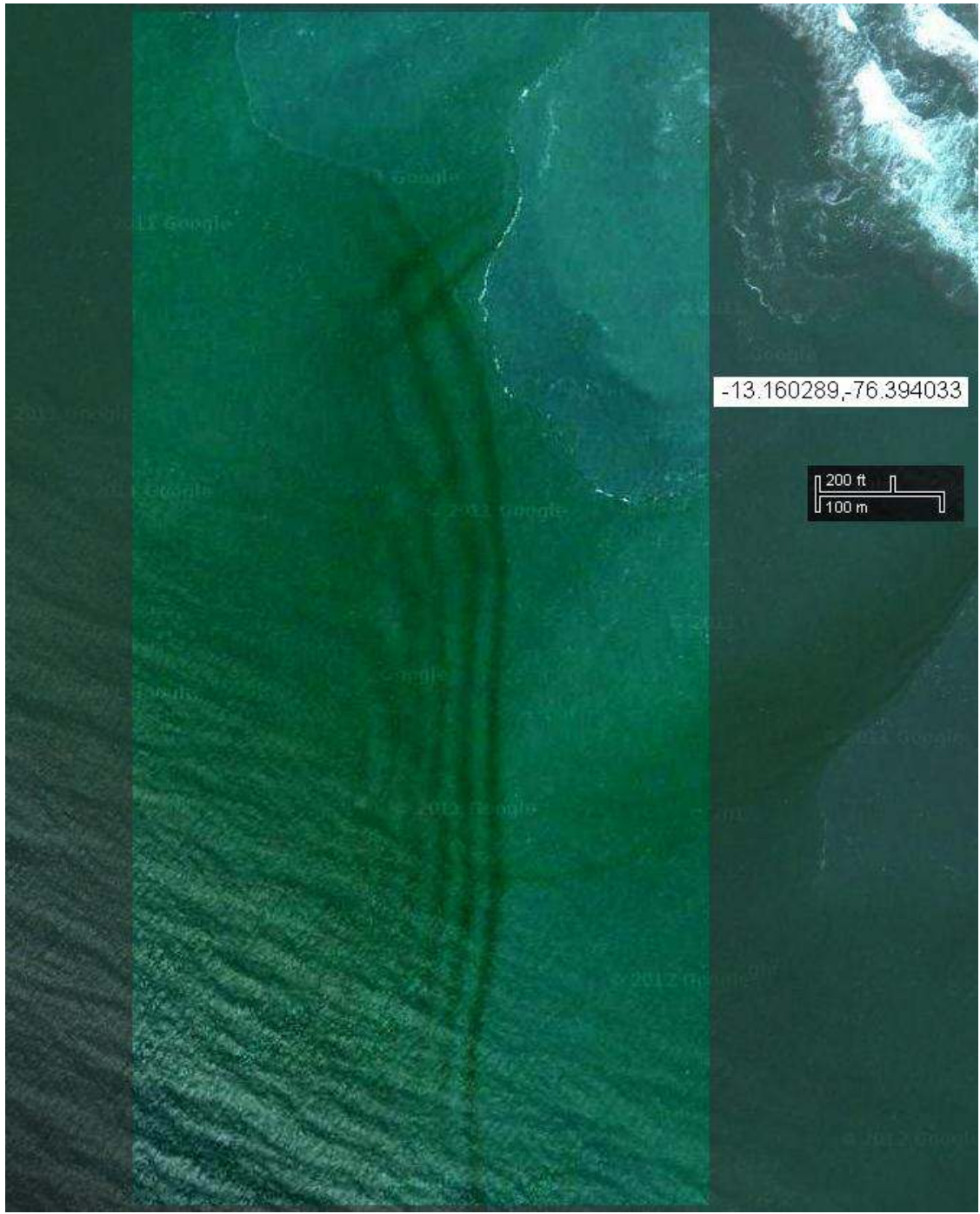

Fig.3 Internal waves near the coastline of Peru.

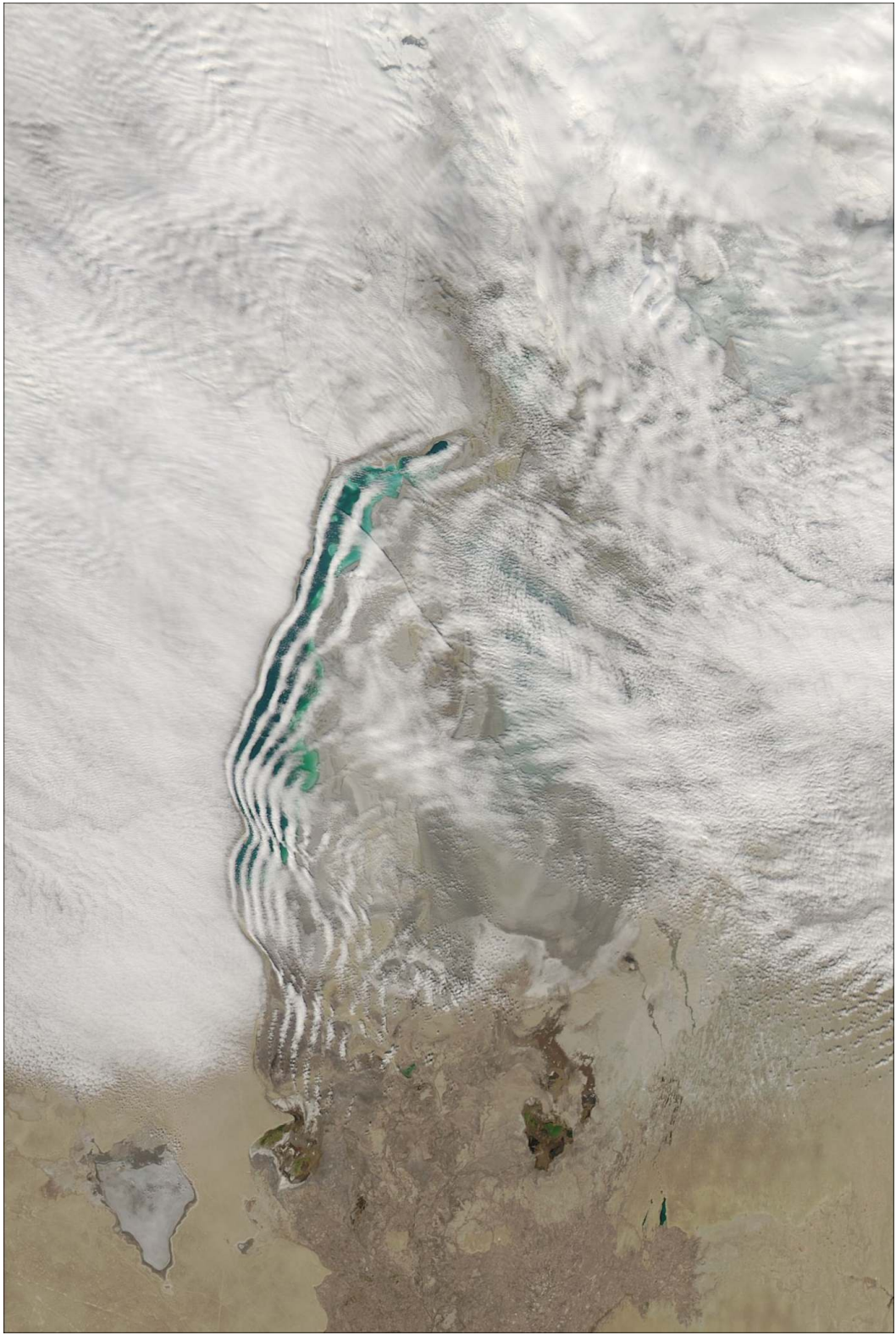

Fig.4 Internal waves in the atmosphere.

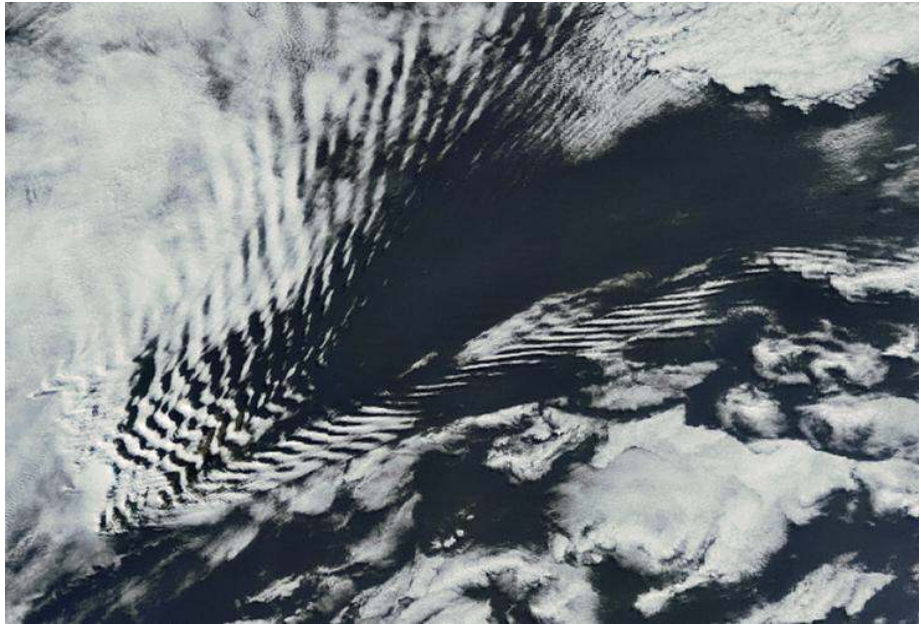

Fig.5 The Kerguelen Islands, also known as Desolation Island, produce the wave clouds, having a form similar to that of the wake created by a ship.

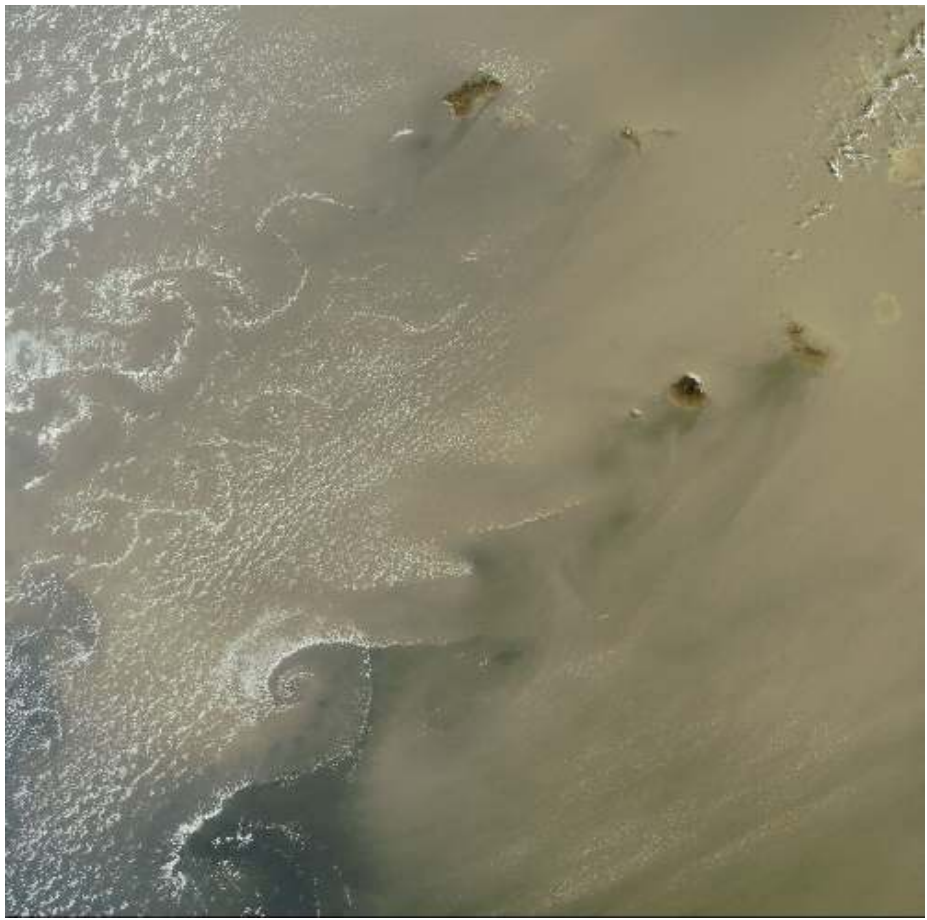

Fig.6 Kármán vortex streets in the dust blowing over Cape Verde Islands.